\documentclass[12pt,a4paper]{article}

\usepackage{amsmath}
\usepackage{amssymb}
\usepackage{epsfig}
\usepackage{graphicx}
\usepackage{cite}

\newcommand{\capdef}{}
\newcommand{\mycaption}[2][\capdef]{\renewcommand{\capdef}{#2}%
       \caption[#1]{{\footnotesize #2}}}
\makeatletter
\renewcommand{\fnum@table}{\textbf{\tablename~\thetable}}
\renewcommand{\fnum@figure}{\textbf{\figurename~\thefigure}}
\makeatother

\newcounter{myenumi}

\renewcommand{\themyenumi}{\roman{myenumi}}
{\end{list}}

\newlength{\myem}
\newcounter{mysubequation}[equation]

\makeatletter
\renewcommand{\section}{\@startsection{section}{1}{0em}{-\baselineskip}%
{\baselineskip}{\normalfont\large\bfseries}}
\renewcommand{\subsection}%
{\@startsection{subsection}{2}{0em}{-0.7\baselineskip}%
{0.7\baselineskip}{\normalfont\bfseries}}
\makeatother

\newcommand{\ie}{{\it i.e.}}

\newcommand{\eg}{{\it e.g.}}

\newcommand{\eq}{Eq.}

\newcommand{\fig}{Figure}
\newcommand{\Fig}{Figure}

\newcommand{\Ref}{Ref.}
\newcommand{\Refs}{Refs.}
\newcommand{\Sec}{Section}
\newcommand{\Secs}{Sections}

\newcommand{\Tab}{Table}

\newcommand{\stheta}{\sin^22\theta_{13}}

\newcommand{\bi}{\begin{itemize}}
\newcommand{\ei}{\end{itemize}}

\begin{document}


\renewcommand{\thefootnote}{\alph{footnote}}

\begin{flushright}
MADPH-04-1401\\
TUM-HEP-565/04\\
SISSA 83/2004/EP\\
\end{flushright}

\vspace*{1.5cm}

\renewcommand{\thefootnote}{\fnsymbol{footnote}}
\setcounter{footnote}{-1}

{\begin{center}
{\Large\textbf{R2D2 -- a symmetric measurement of reactor neutrinos 
free of systematical errors}}
\end{center}}
\renewcommand{\thefootnote}{\it\alph{footnote}}

\vspace*{.8cm}

{\begin{center} {{\bf
                P.~Huber\footnote[1]{\makebox[1.cm]{Email:}
                \sf phuber@pheno.physics.wisc.edu},
                M.~Lindner\footnote[2]{\makebox[1.cm]{Email:}
                \sf lindner@ph.tum.de} and
                T.~Schwetz\footnote[3]{\makebox[1.cm]{Email:}
                \sf schwetz@sissa.it}
                }}
\end{center}}
{\it
\begin{center}
       $^{a}$Department of Physics, University of Wisconsin\\
       1150 University Avenue, Madison, WI 53706, USA\\[4mm]
       $^{b}$Physik--Department, Technische Universit\"at M\"unchen\\
       James--Franck--Strasse, D--85748 Garching, Germany\\[4mm]
       $^c$Scuola Internazionale Superiore di Studi Avanzati\\
       Via Beirut 2--4, I--34014 Trieste, Italy
\end{center}}

\vspace*{0.5cm}

\begin{abstract}
We discuss a symmetric setup for a reactor neutrino oscillation experiment 
consisting of two reactors separated by about 1 km, and two symmetrically
placed detectors, one close to each reactor. We show that such a configuration 
allows a determination of $\sin^22\theta_{13}$ which is essentially free of 
systematical errors, if it is possible to separate the contributions 
of the two reactors in each detector sufficiently. This can be achieved 
either by considering data when in an alternating way only one reactor is 
running or by directional sensitivity obtained from the neutron displacement
in the detector.  
\end{abstract}

\vspace*{.5cm}


\newpage

\renewcommand{\thefootnote}{\arabic{footnote}}
\setcounter{footnote}{0}

\section{Introduction}

In the previous several years large progress has been obtained in
neutrino physics, and quite a clear picture is emerging for the mass
and mixing parameters relevant for neutrino oscillations (see
\Ref~\cite{Maltoni:2004ei} for a recent review). One of the remaining
questions is the value of the mixing angle $\theta_{13}$, where
present data give only the upper bound $\sin^2 2\theta_{13} \le 0.086$
at 90\% CL \cite{Maltoni:2004ei}, and it is a main objective of the
next generation of oscillation experiments to pin down the value of
$\theta_{13}$ (see \Ref~\cite{Huber:2004ug} for an overview and
references). The current bound on $\stheta$ is dominated by the
reactor neutrino experiments CHOOZ~\cite{Apollonio:2002gd} and Palo
Verde~\cite{Boehm:2001ik}, which have been looking for the
disappearance of the electron anti-neutrinos produced by nuclear
reactors with a detector located about $1\,\mathrm{km}$ away from the
reactor core. The sensitivity of these experiments is limited by the
systematical uncertainties related to the neutrino flux produced by
the reactor, which typically are of the order of 2\%. 

Recently it has been realized that the sensitivity of reactor neutrino
experiments can be significantly improved if in addition to the ``far
detector'' a ``near detector'' is put at a distance $\lesssim
300\,\mathrm{m}$ from the
reactor~\cite{Martemyanov:2002td,Minakata:2002jv,Huber:2003pm}. Many
sites around the world are under discussion, among them the
Double-Chooz proposal~\cite{Ardellier:2004ui,Berridge:2004gq} at the
original CHOOZ site in France, the KASKA project in
Japan~\cite{Minakata:2002jv} at the Kashiwazaki-Kariwa reactor
complex, a site in China at the Daya-Bay reactor complex~\cite{wang}
(see also \Ref~\cite{Liu:2004xp}), and several sites in the
US~\cite{Shaevitz:2003ws}, \eg, at the Braidwood reactor (see
\Ref~\cite{Anderson:2004pk} for an overview). With such a
one-reactor-two-detector setup (R1D2) the initial neutrino flux before
oscillations can be determined to a very good accuracy by the near
detector. The uncertainties associated to the neutrino flux are
eliminated by the comparison of the near and far spectra, and
sensitivities of the order of $10^{-2}$ for $\stheta$ can be achieved.
To reach this aim it is necessary to reduce the uncertainties
associated with the detectors below the one percent level, since the
performance depends crucially on the relative normalization of the two
detectors (see \Ref~\cite{Huber:2003pm} for a detailed discussion). 

\begin{figure}[t!]
   \centering \includegraphics[width=0.7\textwidth]{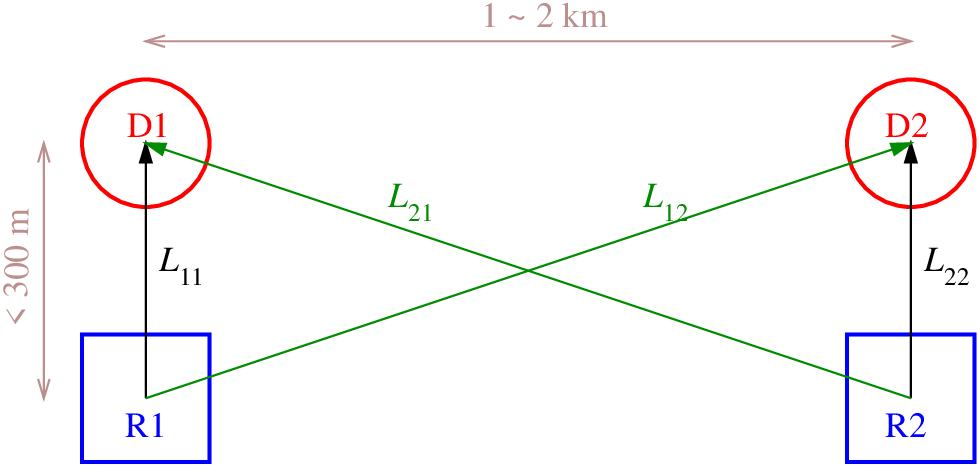}
    \mycaption{Geometry of the symmetric reactor neutrino experiment.
    The two reactor cores are labeled R1 and R2, and the two detectors
    are labeled D1 and D2.} \label{fig:geometry}
\end{figure}

In the present note we discuss a reactor neutrino measurement
consisting of two reactors separated about one kilometer, and two
symmetrically placed detectors, one close to each reactor. This setup
is called R2D2 and it is illustrated in \fig~\ref{fig:geometry}.  Such
configurations could be realized, \eg, at the Kashiwazaki-Kariwa 
complex in Japan or at the Daya-Bay complex in China.\footnote{In the
case of Kashiwazaki-Kariwa there are actually two reactor groups
consisting of 3 and 4 blocks. In this work we always consider the
idealized situation of only two separated reactor cores.} A crucial
element is the assumption that it is possible to separate events from
the two reactors within the detectors. This could be achieved by
periods where successively one of the two reactors is off or with both
reactors on by sufficient statistical directional sensitivity obtained
from the measurement of the neutron 
displacement~\cite{Apollonio:1999jg,Vogel:1999zy} in the detector.  We
will show in the following that the measurement of $\stheta$ becomes
in this case essentially free of systematical uncertainties, and the
$\stheta$ sensitivity is limited only by statistics.  The reason for this is
that all uncertainties associated to the produced neutrino flux cancel
by the comparison of the two detectors, and the uncertainties
associated to the detectors cancel since each detector acts
simultaneously as near and far detector. Therefore, such a setup leads
to a very efficient ``self-calibration'' and it turns out that in the
ideal case indeed {\it no} external information on flux and detector
normalization is necessary.

Let us mention that a similar effect could be achieved in principle at
a one-reactor experiment if near and far detectors are exchanged.
Movable detectors are discussed for some proposals in the US, see
\Refs~\cite{Shaevitz:2003ws,Anderson:2004pk}. However, the R2D2
configuration provides a very elegant way to benefit from such kind of
cross-calibration without the need to move detectors, which might
introduce additional systematical effects. We also note that in a
recent paper~\cite{Sugiyama:2004bv} a general
multi-reactor-multi-detector experiment has been considered, which 
includes in principle also the R2D2 configuration. However, in that
analysis, for the near detector(s) the events induced by far reactors
have been neglected, and therefore the effects discussed in the
present note are not covered.

The outline of the paper is as follows. In \Sec~\ref{sec:analytical}
we present analytical considerations to illustrate how systematic
effects associated to the detectors cancel in the R2D2 setup, whereas
in \Sec~\ref{sec:numerical} we show the results of more realistic
numerical investigations, including several types of systematics and
backgrounds. In \Sec~\ref{sec:separation} we comment on the
possibility to use the neutron displacement to separate the two
reactor contributions and on how the performance of R2D2 is affected
if it is not possible to separate the events from the two reactors
perfectly. We finish the work with concluding remarks in
\Sec~\ref{sec:conclusions}.


\section{Analytical estimates}
\label{sec:analytical}

To illustrate how systematic effects associated to the detectors 
cancel in the R2D2 setup we consider in the following a total rate
measurement, taking into account only the (individual) normalization
errors $\sigma_1 = \sigma_2 \equiv \sigma_\mathrm{det}$ of the two
detectors. It is straight forward to include in this calculation more
systematics like flux uncertainties or energy calibration, which we
omit in this section for the sake of clarity. In this simplified case the
$\chi^2$ to estimate the sensitivity to $\stheta$ is
given by
\begin{equation}\label{eq:chisq1}
\chi^2_\mathrm{R2D2} = \sum_{r,d=1,2} \frac{\left[ (1+a_d)T_{rd} -
O_{rd} \right]^2}{O_{rd}} + \sum_{d = 1,2} \left(
\frac{a_d}{\sigma_\mathrm{det}} \right)^2 \,,
\end{equation}
where $d$ denotes the detector and $r$ the reactor. The theoretical
prediction in the case of oscillations $T_{rd}$ and the ``observed''
number of events $O_{rd}$ are given by
\begin{eqnarray}
T_{rd} &=& N_{rd} 
( 1 - \sin^22\theta_{13} \langle \sin^2\phi\rangle_{rd})
\label{eq:theor}\\
O_{rd} &=& N_{rd} \,,
\end{eqnarray}
where $N_{rd}$ denotes the number of events in detector $d$ from
reactor $r$ without oscillations, and $\langle \sin^2\phi\rangle_{rd}$
is the averaged oscillation phase with $\phi = \Delta m^2_{31} L_{rd}
/ (4E_\nu)$. Up to leading order in the small quantities $a_d$ and
$\sin^22\theta_{13}$ \eq~(\ref{eq:chisq1}) becomes
\begin{equation}\label{eq:chisq2}
\chi^2_\mathrm{R2D2} = \sum_{r,d=1,2} 
N_{rd} 
\left( a_d - \sin^22\theta_{13} \langle \sin^2\phi\rangle_{rd} \right)^2  
+
\sum_{d=1,2} \left( \frac{a_d}{\sigma_\mathrm{det}} \right)^2 \,.
\end{equation}
To minimize \eq~(\ref{eq:chisq2}) with respect to $a_1$ and $a_2$ we
calculate
\begin{equation}
\frac{\partial \chi^2_\mathrm{R2D2}}{\partial a_d} =
2 \sum_{r=1,2} 
N_{rd} 
\left( a_d - \sin^22\theta_{13} \langle \sin^2\phi\rangle_{rd} \right)
+
2\, \frac{a_d}{\sigma_\mathrm{det}^2} = 0\,,
\end{equation}
which gives
\begin{equation}\label{eq:a1}
a_d = \sin^22\theta_{13} \, 
\frac{ \sum_{r=1,2} N_{rd} \langle \sin^2\phi\rangle_{rd} }
     { \sum_{r=1,2} N_{rd} + 1/\sigma_\mathrm{det}^2 } \,.
\end{equation}

For simplicity let us assume complete symmetry between the detectors
and reactors. Then we can identify the indices (compare
\fig~\ref{fig:geometry})
\begin{equation}
\begin{array}{c}
(r,d) = (1, 1) = (2, 2) = N \,,\\
(r,d) = (1, 2) = (2, 1) = F \,,
\end{array}
\end{equation}
and hence, $N_{11} = N_{22} = N_N$ and $N_{12} = N_{21} = N_F$.
Furthermore, we assume no oscillations for the ``near'' baseline, \ie,
$\langle \sin^2\phi\rangle_N = 0$, and the oscillation maximum at the
``far'' baseline \ie, $\langle \sin^2\phi\rangle_F \sim 1$. Then
\eq~(\ref{eq:a1}) becomes
\begin{equation}\label{eq:a2}
a_1 = a_2 \approx \sin^22\theta_{13} \, 
\frac{ N_F \langle \sin^2\phi\rangle_F}
     { N_N + N_F + 1/\sigma_\mathrm{det}^2  } \approx
 \sin^22\theta_{13} \, \langle \sin^2\phi\rangle_F \,
 \frac{ N_F }{ N_N } \,,
\end{equation}
where we have used $N_N \gg N_F + 1/\sigma_\mathrm{det}^2$.  For the
kind of setup under consideration one finds $\langle
\sin^2\phi\rangle_F \, N_F / N_N = \langle \sin^2\phi\rangle_F (L_N /
L_F)^2 \sim 10^{-2}$. Hence, we conclude from \eq~(\ref{eq:a2}) that
the systematic pulls $a_1$ and $a_2$ are very close to zero. Indeed,
substituting \eq~(\ref{eq:a2}) into \eq~(\ref{eq:chisq2}) we find
\begin{equation}
\chi^2_\mathrm{R2D2} = 2 N_F \, 
\left( \sin^22\theta_{13} \, 
\langle \sin^2\phi\rangle_F \right)^2 \,
\left[ 1 - \mathcal{O}\left(\frac{N_F}{N_N}\right) \right] \,,
\end{equation}
which gives the 1$\sigma$ error on $\stheta$ of
\begin{equation}\label{eq:error}
\sigma_\mathrm{R2D2}(\stheta) = 
\frac{1}{\sqrt{2N_F} \, \langle \sin^2\phi\rangle_F}
\left[ 1 + \mathcal{O}\left(\frac{N_F}{N_N}\right) \right] \,.
\end{equation}
The leading order term corresponds to the case $a_1 = a_2 = 0$ in
\eq~(\ref{eq:chisq2}), where the error on $\sin^22\theta_{13}$ is
essentially given by the statistical error $1/\sqrt{2N_F}$. Note that
the result in \eq~(\ref{eq:error}) is independent of the value of
$\sigma_\mathrm{det}$, which means that within this framework {\it no}
external information is needed on the detector normalization. 

Let us now compare this result with the traditional R1D2 setup with
only one reactor.  By dropping the reactor index $r$ and taking into
account a factor of two in the number of events to obtain the same
statistics a similar calculation leads to
\begin{equation}\label{eq:errorR1D2}
\sigma_\mathrm{R1D2}(\stheta) =
\sqrt{\frac{1 + 2N_F \sigma_\mathrm{det}^2}{2N_F}} \,
\frac{1}{\langle \sin^2\phi\rangle_F} \,.
\end{equation}
\eq~(\ref{eq:errorR1D2}) gives the same result as \eq~(\ref{eq:error})
in the statistics dominated regime $N_F \sigma_\mathrm{det}^2 \ll 1$.
However, once the statistical and systematical errors become
comparable, \ie, $1/\sqrt{2N_F} \sim \sigma_\mathrm{det}$ the
1$\sigma$ error on $\stheta$ from the R1D2 setup becomes larger than
the one from R2D2. Eventually, in this simple analysis based only on
the total number of events, for infinite statistics the one-reactor
setup is limited by the systematical error:
$\sigma_\mathrm{R1D2}(\stheta) \to \sigma_\mathrm{det} /\langle
\sin^2\phi\rangle_F$ for $N_F \to \infty$, whereas R2D2 gives infinite
precision: $\sigma_\mathrm{R2D2}(\stheta) \to 0$.


\section{Numerical calculations}
\label{sec:numerical}

In this section we present more realistic numerical calculations to confirm
the analytical estimates of the previous section. We assume two identical
reactors and two identical detectors and adopt typical values for the
baselines: $L_{11} = L_{22} = 300\,\mathrm{m}$ and $L_{12} = L_{21} =
1.3\,\mathrm{km}$ (compare \fig~\ref{fig:geometry}). However, we remark that
minor deviations from such a perfect symmetry do not change the qualitative
behaviour of our results. To demonstrate the maximum potential of the R2D2
configuration we assume in this section a {\it perfect} separation of the
events from the two reactors. We will comment on how deviations from this
idealized situation affect the performance later in \Sec~\ref{sec:separation}.
Following \Refs~\cite{Huber:2003pm,Huber:2004ug,Huber:2004xh} we perform a
$\chi^2$ analysis taking into account full spectral information (15 bins in
positron energy) and various systematical effects, which are listed in
\Tab~\ref{tab:systematics}.\footnote{We do not include the spectral shape
uncertainty of the initial neutrino flux, since it was shown in
\Ref~\cite{Huber:2004xh} that for these type of experiments it has a very
small impact on the sensitivity.}

\begin{table}
\centering
\begin{tabular}{|l|p{0.8\linewidth}|}
\hline\hline 
  Label & Type of systematical effect \\ 
  \hline 
  (S1) & 2\% error on the neutrino flux normalizations (uncorrelated
  between the reactors), 0.6\% error on the detector fiducial masses
  (uncorrelated between the detectors), energy scale uncertainty of
  0.5\% for each detector. \\ 
  (S2) & Four different backgrounds of known shape (in total 2.2\% of
  the events from the far reactor), with a normalization
  uncertainty of 100\% for each component, uncorrelated between the two
  detectors (see \Ref~\cite{Huber:2004ug} for details).\\ 
  (S3) & Flat background (2\% of the events from the far reactor) with
  a bin-to-bin uncorrelated uncertainty of 50\%.\\
  (S4) & Experimental error completely uncorrelated between bins,
  detectors, and reactors.\\ \hline\hline
\end{tabular}
   \mycaption{Systematical effects and backgrounds considered in the
   numerical calculations.} \label{tab:systematics}
\end{table}

\begin{figure}[t!]
   \centering \includegraphics[width=0.9\textwidth]{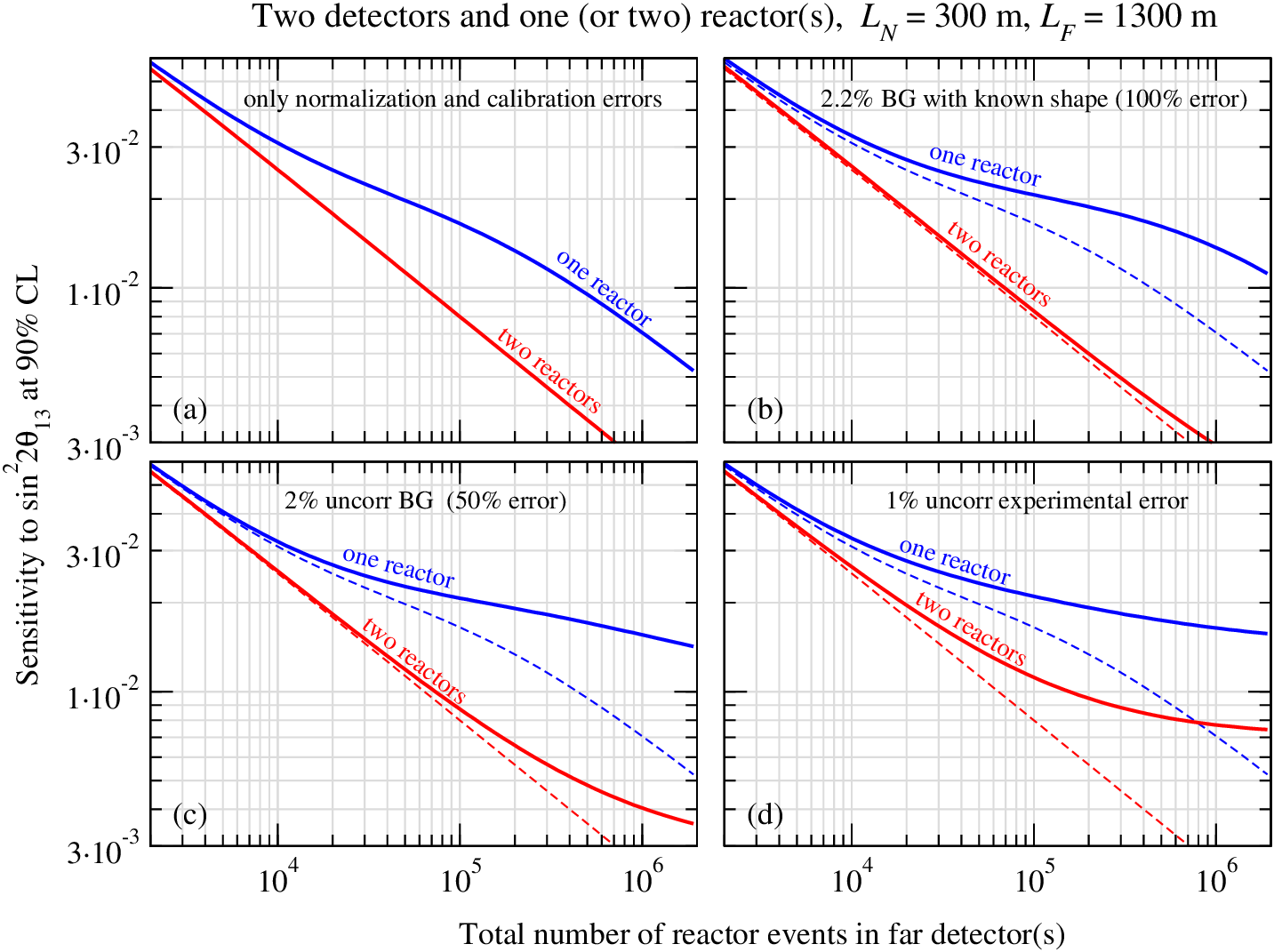}
    \mycaption{Comparison of the symmetric R2D2 reactor neutrino
    experiment setup (``two reactors'') with the R1D2 configuration
    (``one reactor''). The luminosity scaling of the sensitivity to
    $\stheta$ is shown for different assumptions on backgrounds and
    systematic errors as described in \Tab~\ref{tab:systematics}. In
    all panels flux and detector normalization errors, as well as the
    energy scale uncertainty (S1) are included. In panels (b), (c),
    (d) in addition the systematics (S2), (S3), (S4) are included,
    respectively.  The dashed curves in the panels (b), (c), and (d)
    correspond to the curves shown in panel (a).} \label{fig:lumi}
\end{figure}

The results of our analysis are summarized in \fig~\ref{fig:lumi}. We
show the sensitivity to $\stheta$ for the R2D2 setup as a function of
the total number of far detector events, \ie, $N_{12} +
N_{21}$.\footnote{At the baseline of $1.3\,\mathrm{km}$ roughly 130
events are obtained per ton detector mass, GW thermal reactor
power, and years of data taking time.} To compare this luminosity
scaling with the one of the traditional R1D2 setup we consider for the
same configuration only events from one reactor and rescale the event
numbers by a factor of 2 to obtain the same statistics as for R2D2.
Panel (a) of \fig~\ref{fig:lumi} confirms the result of the previous
section: In the case of the R2D2 setup the sensitivity to $\stheta$
scales simply with the square root of the number of events. The
normalization uncertainties of fluxes and detectors cancel thanks to
the symmetric cross-calibration implied by the experimental
configuration. We have verified that even if both reactor fluxes and
both detector normalizations are treated as free parameters in the fit
the result is practically unchanged. In contrast, in the case of R1D2
we observe the well known effect of the worsening of the sensitivity
caused by the detector uncertainties~\cite{Huber:2003pm}.

In panels (b) and (c) of \fig~\ref{fig:lumi} we have investigated the
impact of the backgrounds (S2) and (S3) described in
\Tab~\ref{tab:systematics}. One observes that the sensitivity of the
symmetric reactor experiment is hardly affected by backgrounds,
whereas in the case of the one-reactor configuration they may lead to
a significant worsening of the performance, especially for high
statistics. The reason is again, that for R2D2 the background can
affect a given detector only in the same way for the events from the
near reactor and the far reactor. Hence, it is impossible to mimic the
effect of oscillations by the background. Note that this is also true
if the magnitude and/or the shape of the backgrounds in the two
detectors are different. Only for very high luminosities a bin-to-bin
uncorrelated background (S3) leads to some worsening of the
sensitivity also for the R2D2 configuration.

Finally we have considered the impact of a completely uncorrelated error (S4)
in panel~(d) of \fig~\ref{fig:lumi}. As expected in that case the
$1/\sqrt{N}$-scaling is prevented also for the R2D2 setup.  However, we want
to stress that such an error corresponds to the worst case situation and
realistically one expect such an error to be much less than 1\%.  It is hard
to imagine a physical effect which can lead to an error which is uncorrelated
between the detectors as well as between the reactors.\footnote{Let us note
that an error of this type is expected to be significantly smaller than a
completely uncorrelated error in the case of R1D2, as considered \eg\ in
\Refs~\cite{Huber:2003pm,Ardellier:2004ui}. The reason is that contributions
to this error related to the reactors {\it and} to the detectors are
eliminated because of the self-calibration, similar as in the case of the
errors (S1).}  In the case where the separation of the events from the two
reactors is done only by using data when one of the reactors is shut down time
dependent effects in the detector or the backgrounds could introduce some
uncorrelated errors, since data taken at different times are compared. If the
event separation is done by the neutron displacement measurement a background
varying with the (horizonal) direction could lead to such an error. Both of
these effects are expected to be very small.


\section{Neutron displacement and the statistical separation of
events} \label{sec:separation}

A crucial assumption of the considerations presented in
\Secs~\ref{sec:analytical} and \ref{sec:numerical} is that it is
possible to separate the contributions of the two reactors in the
detectors. This separation can be done in an obvious way by just
considering data only when one of the two reactors is switched off.
However, in that case it will be very difficult to obtain a large
number of events. Therefore we point out a second possibility to
separate the events from simultaneously running reactors using the
directional sensitivity, which is based on the displacement of the
neutrons between production point and capture on the scintillator. 
This effect has been observed previously in 
the reactor experiments G{\"o}sgen~\cite{Zacek:1986cu},
Bugey~\cite{Declais:1994su} and CHOOZ~\cite{Apollonio:1999jg}.

After the production of the neutron in the reaction $\bar\nu_e + p \to
e^+ + n$ it is moderated to thermal energies by scattering on
hydrogen. This is a non-isotropic process which preserves some memory
of the initial direction and leads to an average displacement of about
1.5~cm along the direction of the incoming neutrino. Once the thermal
energy is reached isotropic diffusion takes place before the neutron
capture, which does not alter the average displacement, but leads to a
smearing of the individual capture locations. The details of this
process depend also on the Gd loading of the scintillator. Finally one
obtains a 3-dimensional Gaussian distribution of neutron capture
locations, with a mean shifted from the interaction vertex by
approximately 1.5~cm along the neutrino direction. The width of the
distribution is determined by the smearing from the diffusion process
and the spatial detector resolution, and a typically value for the
width is 6~cm~\cite{oberauer} (see also \Ref~\cite{Vogel:1999zy}).
Although the width of the Gaussian distribution is several times
larger than the signal the large number of events allows a rather
precise determination of the average neutron
displacement.\footnote{Recall that if $N$ events are observed from a
Gaussian distribution with standard deviation $\sigma$ the error on
the mean is given by $\sigma / \sqrt{N}$.} For example with the 2700
events in CHOOZ a precision of $(1.9\pm0.4)$~cm has been obtained for
the neutron displacement~\cite{Apollonio:1999jg}. 

This result suggests that it might be possible to use this effect to
separate the two reactor contributions in the case of the R2D2
experiment. Assuming an angle of $180^\circ$ between the two reactors
it is necessary to identify two overlapping Gaussian distributions,
shifted by $\pm 1.5$~cm and with widths of $\sim 6$~cm. Hence it is
impossible to determine the source reactor for a single event, however
on a statistical basis the number of events from the two reactors
could be determined by fitting the normalizations of the two
distributions to the data. Although there is a large overlap of the
two Gaussian distributions the large number of events imply very small
statistical errors, and it seems possible to separate the two reactor
contributions with sufficient precision. Moreover, also in this case
the accuracy benefits from the cross-calibration offered by the
geometrical configuration. 

The obtainable accuracy in the event separation based on this method
may very well reach the percent level. To illustrate this claim we
assume that the neutron displacement is $1.5\,\mathrm{cm}$ and the
corresponding width of the Gaussian distribution is 
6~cm~\cite{oberauer,Vogel:1999zy}.
%
%
Next we take the distance to the far reactor to be $1.3\,\mathrm{km}$,
whereas the distance to the near reactor is taken to be
$300\,\mathrm{m}$.  The reactors and detectors are aligned such that
the angle between the direction to the near and far reactor in each
detector is $180^\circ$.  Now we simulate data with the correct
angular distributions and try to determine the number of events from
the far reactor in one detector by fitting the simulated data. Based
on a sample of $10^5$ events from the far reactor we find with above
assumptions that the far reactor events can be measured with an
accuracy of 3\% at $1\sigma$.

\begin{figure}[t!]
   \centering \includegraphics[width=0.6\textwidth]{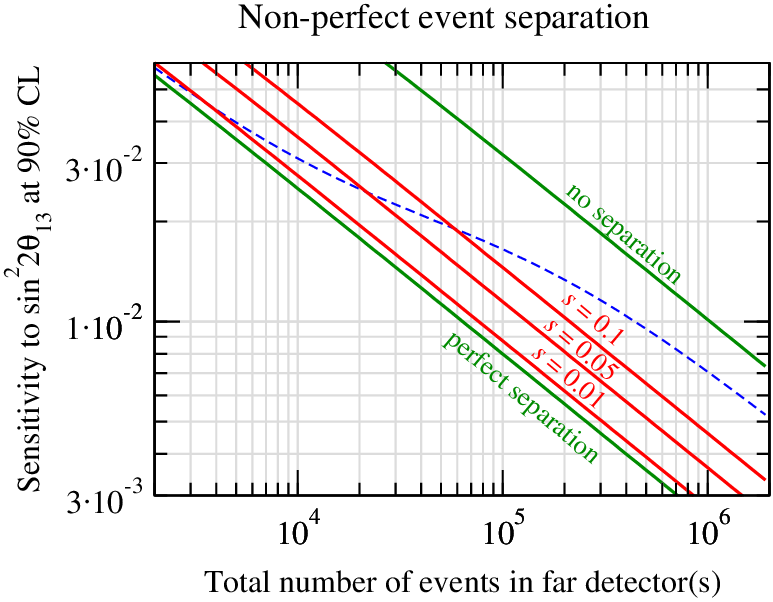}
   \mycaption{Luminosity scaling of the $\stheta$ sensitivity of the
   R2D2 setup in the case of non-perfect separation of the two reactor
   contributions. The same setup as in \fig~\ref{fig:lumi}(a)
   has been adopted. The sensitivity is shown for various values of
   the fraction of miss-identified events $s$ (see
   \eq~(\ref{eq:separation})). The dashed curve corresponds to the
   one-reactor setup R1D2.} \label{fig:separation}
\end{figure}

A detailed investigation of this possibility goes beyond the scope of
the present work and will be covered elsewhere~\cite{newpaper}. In the
following we want to estimate the impact of a non-perfect
separation of the two reactor contributions on the sensitivity to
$\stheta$.
Therefore we define the parameter $s$ as the fraction of 
miss-identified events, which is given from the statistical precision
of a fit to the directional information as outlined above. We perform
the same $\chi^2$ analysis as in \Sec~\ref{sec:numerical} but we
replace the event numbers
\begin{equation}\label{eq:separation}
\begin{array}{l@{\:\to\:}l}
N_{1d} & (1-s)N_{1d} + sN_{2d} \,,\\
N_{2d} & (1-s)N_{2d} + sN_{1d} 
\end{array}
\end{equation}
in each energy bin for the two detectors $d = 1,2$. Hence, $s=0$ (as
well as $s=1$) corresponds to perfect separation and the results of 
\Sec~\ref{sec:numerical} are recovered, whereas $s=0.5$ corresponds to
the case where no distinction of the reactors is possible.  In
\fig~\ref{fig:separation} we compare the luminosity scaling of the
$\stheta$ sensitivity of the R2D2 setup for various values of $s$. It
becomes clear that one has to achieve $s \lesssim 0.05$ in order to
maintain the sensitivity. Note however, that for luminosities $\gtrsim
5\times 10^4$ R2D2 is still better than the one-reactor setup R1D2 even
for $s=0.1$. 

In order to understand these results we note that in the case of
non-perfect event separation the spectral analysis is very important. 
The energy distortion due to oscillations is present only in the
events from the far reactor, and hence it provides a unique signal,
which leads to a non-trivial sensitivity to $\stheta$ even if no
distinction of the two reactors is possible (see
\Fig~\ref{fig:separation}). Consequently the results shown in 
\Fig~\ref{fig:separation} depend to some extent on the details of the
spectral analysis, \eg, number of bins or energy resolution. We add
however, that the main conclusions drawn from this figure are stable,
and do not change also in the presence of additional systematical
effects and/or backgrounds. The investigation of higher order effects
like an energy dependent event separation $s$ or an uncertainty on $s$
is beyond the scope of the present note.


\section{Concluding remarks}
\label{sec:conclusions}

In this note we have pointed out that a reactor neutrino measurement
with a configuration consisting of two separated reactors and two 
symmetrically placed nearby detectors (R2D2) allows a determination of
$\stheta$ essentially free of systematical errors. The cross-calibration 
which becomes possible in such a configuration eliminates uncertainties 
associated to the initial neutrino fluxes as well as detector effects. 
In addition the impact of backgrounds is very small. 
In the specific cases of the KASKA proposal in Japan or the Daya-Bay 
project in China a configuration similar to the R2D2 setup seems in
principle possible, since these reactor complexes consist of two groups 
of cores, separated by a distance of the order of
$1\,\mathrm{km}$, and a near detector has to be located close to each
of the two blocks. Hence, in addition to the data from the far
detector (``far'' from all the reactors) also the data from the two
near detectors on their own could provide very valuable information on
$\stheta$, with a very small impact of systematical uncertainties.

An important condition for the cancellation of errors is the ability 
to separate events from the two reactors within each detector. 
An obvious way to achieve this would be a scenario, where only one
reactor would be running in an alternating way at a given time. 
This is probably not realistic for commercial power reactors, but
it might be realized in cases where a new reactor complex is added 
to an existing one at the right distance. The two detectors could
measure initially with the first reactor on. Later, when the second
complex is turned on, a (shorter) period where the first complex is 
off could be used to obtain the cross-calibration. 

We are aware of the fact that it might be difficult to achieve large
statistics data samples for an R2D2 experiment if data taking is
limited to the periods when only one reactor is running. We discussed
therefore the possibility to separate the events from two
simultaneously running reactors by directional sensitivity from the
measurement of the neutron displacement. Scaling the results obtained
in the CHOOZ experiment to the much higher event numbers of the setups
under consideration we conclude that such a separation might be
possible on a statistical basis. We have estimated that the
distinction of the events has to be better than 1\% to 5\% in order to
maintain the self-calibration. 

We would also like to point out that our results do not rely on a {\it
perfect} symmetrical configuration. The essential point is that both
detectors act simultaneously as near and far detectors, while a certain
stability exists against asymmetries in the size of the reactors and
detectors, in baselines, and in backgrounds. This implies in particular
that one needs not rely on identical reactors and the same overburden for
the two detectors.

In conclusion, we have demonstrated that the method discussed here has
the potential to significantly reduce the impact of systematical
uncertainties for future reactor neutrino experiments. We believe that
it is worth to pursue further studies to establish its experimental
feasibility, including a more detailed investigation of the potential
of the directional sensitivity of the detectors.

\subsection*{Acknowledgement}

We want to thank L.~Oberauer and H.~de~Kerret for information and
discussions on directional sensitivity from the measurement of the
neutron displacement. This study has been supported by the
``Sonderforschungsbereich 375 f{\"u}r Astro-Teilchenphysik der
Deutschen For\-schungs\-ge\-mein\-schaft''. T.S.\ is supported by a
Marie Curie Intra-European Fellowship within the 6th European
Community Framework Programme.


\end{document}